\documentclass{article}%
\usepackage{amssymb}
\usepackage{amsmath}
\usepackage{amsfonts}
\usepackage{graphicx}%
\providecommand{\U}[1]{\protect\rule{.1in}{.1in}}
\newtheorem{theorem}{Theorem}

\newtheorem{corollary}[theorem]{Corollary}

\newtheorem{definition}[theorem]{Definition}

\newtheorem{lemma}[theorem]{Lemma}

\newtheorem{proposition}[theorem]{Proposition}
\newtheorem{remark}[theorem]{Remark}

\begin{document}

\title{Weil-Petersson Volumes of the Moduli Spaces of CY Manifolds }
\author{Andrey Todorov\\University of California,\\Department of Mathematics\\Santa Cruz, CA 95064\\Bulgarian Academy of Sciences,\\Institute of Mathematics\\ul. Acad. Georgy Bonchev No 8\\Sofia 1113, Bulgaria}
\date{March 12, 2006}
\maketitle

\begin{abstract}
In this paper it is proved that the volumes of the moduli spaces of polarized
Calabi-Yau manifolds with respect to Weil-Petersson metrics are rational
numbers. Mumford introduce the notion of a good metric on vector bundle over a
quasi-projective variety in \cite{Mu}. He proved that the Chern forms of good
metrics define classes of cohomology with integer\ coefficients on the
compactified quasi-projective varieties by adding a divisor with normal
crossings. Viehweg proved that the moduli space of CY manifolds is a
quasi-projective variety. The proof that the volume of the moduli space of
polarized CY manifolds are rational number is based on the facts that the
$L^{2}$ norm on the dualizing line bundle over the moduli space of polarized
CY manifolds is a good metric. The Weil-Petersson metric is minus the Chern
form of the $L^{2}$ metric on the dualizing line bundle. This fact implies
that the volumes of Weil-Petersson metric are rational numbers. Also we get
that the Weil-Petersson metric is a good metric. Therefore all the Chern forms
define integer classes of cohomologies.

\end{abstract}
\tableofcontents

\section{Introduction}

\subsection{General Remarks}

There are several metrics naturally defined on the moduli space of Riemann
surfaces. One of them is the Weil-Petersson metric. The Weil-Petersson metric
is defined because of the existence of a metric with a constant curvature on
the Riemann surface. Its curvature properties were studied by Ahlfors, Bers,
S. Wolpert and so on.

The generalization of the Weil-Petersson metric on the moduli space of higher
dimensional projective varieties was first introduced by Y.-T. Siu. He gave
explicit formulas for the curvature of Weil-Petersson metric. See \cite{Siu}.
The generalization is possible thanks to the solution of Calabi conjecture due
to Yau. See \cite{Yau}. For Calabi-Yau manifolds it was noticed in \cite{To89}
and \cite{Ti} that the Weil-Petersson metric can be defined and computed by
using the cup product of $(n-1,n)$ forms.

Another metric naturally defined on the moduli space of polarized CY manifolds
is the Hodge metric. The holomorphic sectional curvature of the Hodge metric
is negative and bounded away from zero. The holomorphic curvature of the
Weil-Petersson metric is not negative. Recently some important results about
the relations between the Weil-Petersson metric and Hodge metric were
obtained. See \cite{FL}, \cite{Lu}, \cite{LS} and \cite{LS1}.

Ph. Candelas and G. Moore asked if the Weil-Petersson volumes are finite. For
the importance and the physical interpretation of the finiteness of the
Weil-Petersson volumes to string theory see \cite{DL}, \cite{HM} and \cite{V}.
In this paper we will answer Candelas-Moore question. Moreover we will prove
that the Weil-Petersson volumes are rational numbers. I was informed by Prof.
Lu that he and Professor Sun also proved the rationality of the volumes. See
\cite{LS1}.

In 1976 D. Mumford introduced the notion of good metrics on vector bundles on
quasi-projective varieties in \cite{Mu}. He proved that the Chern forms of
good metrics define classes of cohomology with integer\ coefficients on the
compactified quasi-projective varieties by adding a divisor with normal
crossings. Viehweg proved that the moduli space of prioritized CY manifolds is
a quasi-projective variety. The idea of this paper is to apply the results of
Mumford to the moduli space of CY manifolds. We proved that the $\mathbf{L}%
^{2}$ metric on the dualizing sheaf is good. It was proved in \cite{To89} that
the Chern form of the $\mathbf{L}^{2}$ metric on the dualizing sheaf defines
the Weil-Petersson metric on the moduli space. See also \cite{Ti}. So if we
prove that the $\mathbf{L}^{2}$ metric on the dualizing sheaf is good then it
will imply that the Weil-Petersson volumes are rational numbers. We will
explain what is the meaning of a metric on a line bundle is good one.

According to \cite{W} the moduli space of polarized CY
manifolds$\ \mathfrak{M}_{L}$(M) is a quasi-projective variety. Let
$\overline{\mathfrak{M}_{L}(\text{M})}$ be some projective compactification of
$\mathfrak{M}_{L}$(M) such that
\[
\mathfrak{D}=\overline{\mathfrak{M}_{L}(\text{M})}-\mathfrak{M}_{L}(\text{M})
\]
is a divisor of normal crossings. The meaning that the metric $h$ on a line
bundle over $\mathfrak{M}_{L}($M$)$ is good is the following; Let
$\tau_{\infty}\in\mathfrak{D},$ let $D^{N}$ be an open polydisk containing
$\tau_{\infty},$ then $h$ is a good metric on some line bundle $\mathcal{L}$
defined on $\mathfrak{M}_{L}($M$)$ if the curvature form of the metric of the
line bundle around open sets
\[
D^{N}-D^{N}\cap\mathfrak{D}=\left(  D^{\ast}\right)  ^{k}\times D^{N-k}%
\]
is bounded from above by the Poincare metric on $\left(  D^{\ast}\right)
^{k}$ plus the standard metric on $D^{N-k}.$ This implies that if we integrate
the maximal power of the curvature form over $\mathfrak{M}_{L}($M$)$ we get a
finite number. Moreover such curvature forms are forms with coefficients
distribution in the sense of Schwarz and they define classes of cohomology of
$H^{2}\left(  \overline{\mathfrak{M}_{L}(\text{M})},\mathbb{Z}\right)  .$

Our proof that the $L^{2}$ metric $h$ on the relative dualizing sheaf is a
good metric is based on the construction of a canonical family of holomorphic
forms $\omega_{\tau}$ on the Kuranishi space given in \cite{To89}. The
canonical family of holomorphic forms defines a special holomorphic local
coordinates in the Kuranishi space where the components of the Weil-Petersson
metric are given by
\[
g_{i,\overline{j}}=\delta_{i,\overline{j}}+\frac{1}{6}R_{i,\overline
{j},k,\overline{l}}\tau^{k}\overline{\tau^{l}}+...
\]
Since
\[
h(\omega_{\tau})=\left\Vert \omega_{\tau}\right\Vert ^{2}=(-1)^{\frac
{n(n-1)}{2}}\left(  -\sqrt{-1}\right)  ^{n}%
{\displaystyle\int\limits_{\text{M}}}
\omega_{\tau}\wedge\overline{\omega_{\tau}}%
\]
then around a point
\[
\tau_{\infty}\in\mathfrak{D}=\overline{\mathfrak{M}_{L}(\text{M}%
)}-\mathfrak{M}_{L}(\text{M})
\]
we can compute explicitly $h.$ Let $D^{N}$ be any polydisk in $\overline
{\mathfrak{M}_{L}(\text{M})}$ containing $\tau_{\infty}.$ Let
\[
D^{N}-D^{N}\cap\mathfrak{D}=\left(  D^{\ast}\right)  ^{k}\times D^{N-k}.
\]
Let $\pi:$\textrm{U}$_{1}\times...\times\mathrm{U}_{k}\rightarrow\left(
D^{\ast}\right)  ^{k}$ be the uniformization map. From the results in
\cite{To89} we deduce the following explicit formula for the pull back of the
$L^{2}$ metric on the relative dualizing sheaf $h$ on \textrm{U}$_{1}%
\times...\times\mathrm{U}_{k}\times D^{N-k}$:%
\[
\left\Vert \omega_{\tau}\right\Vert ^{2}|_{\left(  D^{\ast}\right)  ^{k}\times
D^{N-k}}:=h|_{\left(  D^{\ast}\right)  ^{k}\times D^{N-k}}=
\]
$\ $%
\begin{equation}
\left\Vert \omega_{\tau}\right\Vert ^{2}:=h(\tau,\overline{\tau}):=%
{\displaystyle\sum\limits_{i=1}^{k}}
\left(  1-|\tau^{i}|^{2}\right)  +%
{\displaystyle\sum\limits_{j=k+1}^{N}}
\left(  1-|t^{j}|^{2}\right)  +\phi(\tau,\overline{\tau})+\Psi(t,\overline
{t}), \label{I}%
\end{equation}
for $0\leq|\tau^{i}|<1,$ $0\leq\left\vert t^{j}\right\vert <1$ where
$\phi(\tau,\overline{\tau})$ and $\Psi(t,\overline{t})$ are bounded real
analytic functions on the unit disk. The expression $\left(  \ref{I}\right)  $
shows that the $L^{2}$ metric $h$ is a good metric. This implies that the
volumes of Weil-Petersson metrics are finite and they are rational numbers.
Moreover it implies that Weil-Petersson metric is a good metric. So the Chern
forms of it define classes of cohomologies in $H^{2k}\left(  \overline
{\mathfrak{M}_{L}(\text{M})},\mathbb{Z}\right)  $ according to \cite{Mu}$.$

\subsection{Description of the Content of the Paper}

Next we are going to describe the content of each of the \textbf{Sections }in
this article.

In \textbf{Section 2} we review the basis results from \cite{To89} and in
\cite{Ti} about local deformation theory of CY manifolds. We also review the
results of \cite{LTYZ} about the global deformation Theory.

In \textbf{Section 3} we review Mumford Theory of good metrics with
logarithmic growth on vector bundles over quasi-projective varieties developed
in \cite{Mu}.

In \textbf{Section 4} we prove that the $L^{2}$ metric on the dualizing line
bundle over the moduli space is a good metric in the sense of Mumford. This
results implies that the Weil-Petersson volumes are rational numbers.

\subsection{Acknowledgements}

Part of this paper was finished during my visit to MPI Bonn. I want to thank
Professor Yu. I. Manin for his help. Special thanks to Ph. Candelas and G.
Moore for drawing my attention to the problem of the finiteness of the
Weil-Petersson volumes. I want to thank G. Moore for useful conversations long
time ago on this topic and his useful comments.

\section{Moduli of Polarized CY Manifolds}

\subsection{Local Moduli}

Let M be an even dimensional C$^{\infty}$ manifold. We will say that M has an
almost complex structure if there exists a section $I\in C^{\infty}%
($M$,Hom(T^{\ast},T^{\ast})$ such that $I^{2}=-id.$ $T$ is the tangent bundle
\ and $T^{\ast}$ is the cotangent bundle on M. This definition is equivalent
to the following one: Let M be an even dimensional C$^{\infty}$ manifold.
Suppose that there exists a global splitting of the complexified cotangent
bundle $T^{\ast}\otimes\mathbf{C}=\Omega^{1,0}\oplus\Omega^{0,1},$ where
$\Omega^{0,1}=\overline{\Omega^{1,0}}.$ Then we will say that M has an almost
complex structure. We will say that an almost complex structure is an
integrable one, if for each point $x\in$M there exists an open set $U\subset$M
such that we can find local coordinates $z^{1},..,z^{n},$ such that
$dz^{1},..,dz^{n}$ \ are linearly independent in each point $m\in U$ and they
generate $\Omega^{1,0}|_{U}.$

\begin{definition}
\label{belt}Let M be a complex manifold. Let $\phi\in\Gamma(M,Hom(\Omega
^{1,0},\Omega^{0,1})),$ then we will call $\phi$ a Beltrami differential.
\end{definition}

Since $\Gamma($M$,Hom(\Omega^{1,0},\Omega^{0,1}))\backsimeq\Gamma($%
M$,\Omega^{0,1}\otimes T^{1,0}),$ we deduce that locally $\phi$ can be written
as follows: $\phi|_{U}=\sum\phi_{\overline{\alpha}}^{\beta}\overline
{dz}^{\alpha}\otimes\frac{\partial}{\partial z^{\beta}}.$ From now on we will
denote by $A_{\phi}$ the following linear operator:
\[
A_{\phi}=\left(
\begin{array}
[c]{cc}%
id & \phi(\tau)\\
\overline{\phi(\tau)} & id
\end{array}
\right)  .
\]
We will consider only those Beltrami differentials $\phi$ such that
$det$($A_{\phi})\neq0.$ The Beltrami differential \ $\phi$\ defines an
integrable complex structure on M if and only if the following equation
holds:
\begin{equation}
\overline{\partial}\phi=\frac{1}{2}\left[  \phi,\phi\right]  , \label{0}%
\end{equation}
where%
\[
\left[  \phi,\phi\right]  |_{U}:=
\]%
\begin{equation}
\sum_{\nu=1}^{n}\sum_{1\leqq\alpha<\beta\leqq n}\left(  \sum_{\mu=1}%
^{n}\left(  \phi_{\overline{\alpha}}^{\mu}\left(  \partial_{\mu}%
\phi_{\overline{\beta}}^{\nu}\right)  -\phi_{\overline{\beta}}^{\mu}\left(
\partial_{\mu}\phi_{\overline{\alpha}}^{\nu}\right)  \right)  \right)
\overline{dz}^{\alpha}\wedge\overline{dz}^{\beta}\otimes\frac{\partial
}{dz^{\nu}} \label{eq:def}%
\end{equation}
(See \cite{KM}.) Kuranishi proved the following Theorem:

\begin{theorem}
\label{Kur}Let $\left\{  \phi_{i}\right\}  $ be a basis of harmonic $(0,1)$
forms of $\mathbb{H}^{1}($M$,T^{1,0})$ on a Hermitian manifold M. Let $G$ be
the Green operator and let $\phi(\tau^{1},..,\tau^{N})$ be defined as
follows:\
\begin{equation}
\phi(\tau)=\sum_{i=1}^{N}\phi_{i}\tau^{i}+\frac{1}{2}\overline{\partial}%
^{\ast}G[\phi(\tau^{1},...,\tau^{N}),\phi(\tau^{1},...,\tau^{N})].
\label{eq:bel}%
\end{equation}
\textit{There exists }$\varepsilon>0$\textit{\ such that for }$\tau=(\tau
^{1},...,\tau^{N})$ \textit{such that }$|\tau_{i}|<\varepsilon$\textit{\ the
tensor }$\phi(\tau^{1},...,\tau^{N})$\textit{\ is a\ global }$C^{\infty}%
$\textit{\ section of the bundle }$\Omega^{(0,1)}\otimes T^{1,0}$.(See
\cite{KM}.)
\end{theorem}

\subsection{Affine Flat coordinates in the Kuranishi Space}

Based on Theorem \ref{Kur}, the following Theorem is proved in \cite{To89}:

\begin{theorem}
\label{tod1}Let M be a CY manifold and let $\left\{  \phi_{i}\right\}  $ be a
basis of harmonic $(0,1)$ forms with coefficients in $T^{1,0}.$ Then the
equation $\left(  \ref{0}\right)  $ has a solution in the form:%
\[
\phi(\tau)=\sum_{i=1}^{N}\phi_{i}\tau^{i}+\sum_{|I_{N}|\geqq2}\phi_{I_{N}}%
\tau^{I_{N}}=
\]
\begin{equation}
\sum_{i=1}^{N}\phi_{i}\tau^{i}+\frac{1}{2}\overline{\partial}^{\ast}%
G[\phi(\tau^{1},...,\tau^{N}),\phi(\tau^{1},...,\tau^{N})] \label{BD}%
\end{equation}
and $\overline{\partial}^{\ast}\phi(\tau^{1},...,\tau^{N})=0,$ $\phi_{I_{N}%
}\lrcorner\omega_{\text{M}}=\partial\psi_{I_{N}}$ \textit{where }
$I_{N}=(i_{1},...,i_{N})$\ \ \textit{is a multi-index}, \
\[
\phi_{I_{N}}\in C^{\infty}(\text{M},\Omega^{0,1}\otimes T^{1,0}),\tau^{I_{N}%
}=(\tau^{1})^{i_{1}}...(\tau^{N})^{i_{N}}%
\]
\textit{and there exists }$\varepsilon>0$ such that when $|\tau^{i}%
|<\varepsilon$\textit{\ } $\phi(\tau)\in C^{\infty}($M$,\Omega^{0,1}\otimes
T^{1,0})$ \textit{where} $i=1,...,N.$
\end{theorem}

\begin{definition}
\label{flat}Theorem \ref{tod1} implies that the Kuranishi space $\mathcal{K}$
is defined as follows: Let $\varepsilon>0$ be such that the Beltarmi
differentials $\phi(\tau)$ defined by $\left(  \ref{BD}\right)  $ satisfy
Theorem \ref{Kur}, then
\[
\mathcal{K}:\{\tau=(\tau^{1},...,\tau^{N})||\tau^{i}|<\varepsilon\}.
\]
Thus $\tau=(\tau^{1},...,\tau^{N})$ such that $|\tau^{i}|<\varepsilon$ is a
local coordinate system in $\mathcal{K}.$ It will be called the flat
coordinate system in $\mathcal{K}$.
\end{definition}

It is a standard fact from Kodaira-Spencer-Kuranishi deformation theory that
for each $\tau=(\tau^{1},...,\tau^{N})\in\mathcal{K}$ as in Theorem \ref{tod1}
the Beltrami differential $\phi(\tau^{1},...,\tau^{N})$ defines a new
integrable complex structure on M. This means that the points of
$\mathcal{K},$ where
\[
\mathcal{K}:\{\tau=(\tau^{1},...,\tau^{N})||\tau^{i}|<\varepsilon\}
\]
defines a family of operators $\overline{\partial}_{\tau}$ on the $C^{\infty}$
family $\mathcal{K}\times M\rightarrow M$ and $\overline{\partial}_{\tau}$ are
integrable in the sense of Newlander-Nirenberg. Moreover it was proved by
Kodaira, Spencer and Kuranishi that we get a complex analytic family of CY
manifolds $\pi:\mathcal{X\rightarrow K},$ where as $C^{\infty}$ manifold
$\mathcal{X\backsimeq K}\times$M$.$ The family
\begin{equation}
\pi:\mathcal{X\rightarrow K} \label{kur}%
\end{equation}
is called the Kuranishi family. The operators $\overline{\partial}_{\tau}$ are
defined as follows:

\begin{definition}
\label{tod3}Let $\{\mathcal{U}_{i}\}$ be an open covering of M, with local
coordinate system $\{z_{i}^{k}\}$ where $k=1,...,\dim_{\mathbb{C}}$M$=n.$ We
know that the Beltrami differential is given by:
\[
\phi(\tau)=\sum_{j,k=1}^{n}(\phi(\tau^{1},...,\tau^{N}))_{\overline{j}}%
^{k}\text{ }d\overline{z}^{j}\otimes\frac{\partial}{\partial z^{k}}.
\]
\textit{Then it defines the }$\overline{\partial_{\phi}}$ operator associated
with the new complex structure as follows:\textit{ \ }\
\begin{equation}
\left(  \overline{\partial_{\phi}}\right)  _{\tau,\overline{j}}=\frac
{\overline{\partial}}{\overline{\partial z^{j}}}-\sum_{k=1}^{n}(\phi(\tau
^{1},...,\tau^{N}))_{\overline{j}}^{k}\frac{\partial}{\partial z^{k}}.
\label{eq:dibar}%
\end{equation}

\end{definition}

In \cite{To89} the following Theorems were proved:

\begin{theorem}
\label{formsa}There exists a family of holomorphic forms $\omega_{\tau}$ of
the Kuranishi family $\left(  \ref{kur}\right)  $ such that in the coordinates
$(\tau^{1},...,\tau^{N})$ we have%
\begin{equation}
\omega_{\tau}=\omega_{0}-%
{\displaystyle\sum\limits_{i,j}}
\left(  \omega_{0}\lrcorner\phi_{i}\right)  \tau^{i}+%
{\displaystyle\sum\limits_{i,j}}
\omega_{0}\lrcorner\left(  \phi_{i}\wedge\phi_{k}\right)  \tau^{i}\tau
^{j}+O(3). \label{formh}%
\end{equation}

\end{theorem}

\begin{theorem}
\label{forms}There exists a family of holomorphic forms $\omega_{\tau}$ of the
Kuranishi family $\left(  \ref{kur}\right)  $ such that in the coordinates
$(\tau^{1},...,\tau^{N})$ we have%
\[
\left\langle \lbrack\omega_{\tau}],[\omega_{\tau}]\right\rangle =(-1)^{\frac
{n(n-1)}{2}}\left(  \sqrt{-1}\right)  ^{n}%
{\displaystyle\int\limits_{\text{M}}}
\omega_{\tau}\wedge\overline{\omega_{\tau}}=
\]%
\[
1-%
{\displaystyle\sum\limits_{i,j}}
\left\langle \omega_{0}\lrcorner\phi_{i},\omega_{0}\lrcorner\phi
_{j}\right\rangle \tau^{i}\overline{\tau^{j}}+
\]%
\[%
{\displaystyle\sum\limits_{i,j}}
\left\langle \omega_{0}\lrcorner\left(  \phi_{i}\wedge\phi_{k}\right)
,\omega_{0}\lrcorner\left(  \phi_{j}\wedge\phi_{l}\right)  \right\rangle
\tau^{i}\overline{\tau^{j}}\tau^{k}\overline{\tau^{l}}+O(\tau^{5})=
\]%
\[
1-%
{\displaystyle\sum\limits_{i,j}}
\tau^{i}\overline{\tau^{j}}+%
{\displaystyle\sum\limits_{i,j}}
\left\langle \omega_{0}\lrcorner\left(  \phi_{i}\wedge\phi_{k}\right)
,\omega_{0}\lrcorner\left(  \phi_{j}\wedge\phi_{l}\right)  \right\rangle
\tau^{i}\overline{\tau^{j}}\tau^{k}\overline{\tau^{l}}+O(\tau^{5})\text{ }%
\]
and%
\begin{equation}
\left\langle \lbrack\omega_{\tau}],[\omega_{\tau}]\right\rangle \leq
\left\langle \lbrack\omega_{0}],[\omega_{0}]\right\rangle . \label{form}%
\end{equation}

\end{theorem}

\subsection{Weil-Petersson Metric}

It is a well known fact from Kodaira-Spencer-Kuranishi theory that the tangent
space $T_{\tau,\mathcal{K}\text{ }}$at a point $\tau\in\mathcal{K}$ can be
identified with the space of harmonic (0,1) forms with values in the
holomorphic vector fields $\mathbb{H}^{1}($M$_{\tau},T$). We will view each
element $\phi\in\mathbb{H}^{1}($M$_{\tau},T$) as a point wise linear map from
$\Omega_{\text{M}_{\tau}}^{(1,0)}$ to $\Omega_{\text{M}_{\tau}}^{(0,1)}.$
Given $\phi_{1}$ and $\phi_{2}\in\mathbb{H}^{1}(M_{\tau},T),$ the trace of the
map
\[
\phi_{1}\circ\overline{\phi_{2}}:\Omega_{\text{M}_{\tau}}^{(0,1)}%
\rightarrow\Omega_{\text{M}_{\tau}}^{(0,1)}%
\]
at the point $m\in$M$_{\tau}$ with respect to the metric g is simply given
by:\textit{\ \ }\
\begin{equation}
Tr(\phi_{1}\circ\overline{\phi_{2}})(m)=\sum_{k,l,m=1}^{n}(\phi_{1}%
)_{\overline{l}}^{k}(\overline{\phi_{2})_{\overline{k}}^{m}}g^{\overline{l}%
,k}g_{k,\overline{m}} \label{wp}%
\end{equation}

\begin{definition}
\label{WP}We will define the Weil-Petersson metric on $\mathcal{K}$ via the
scalar product:\textit{\ \ }\
\begin{equation}
\left\langle \phi_{1},\phi_{2}\right\rangle =\int\limits_{\text{M}}Tr(\phi
_{1}\circ\overline{\phi_{2}})vol(g). \label{wp1}%
\end{equation}

\end{definition}

A very natural construction of a coordinate system $\tau=(\tau^{1}%
,...,\tau^{N})$ in $\mathcal{K}$ is constructed in \cite{To89} such that the
components $g_{i,\overline{j}}$ of the Weil Petersson metric are given by the
following formulas:%

\[
g_{i,\overline{j}}=\delta_{i,\overline{j}}+\frac{1}{6}R_{i,\overline
{j},l,\overline{k}}\tau^{l}\overline{\tau^{k}}+O(\tau^{3}).
\]
\footnote{This coordinate system is called flat holomorphic coordinate system.
It appeared for the first time in \cite{To89}. Based on the information of the
author of \cite{Ti}, it is claimed in \cite{BCOV} that the flat coordinate
system was introduced in \cite{Ti}. The problem of the construction of the
flat holomorphic coordinates was not addressed in \cite{Ti}.}

Very detailed treatment of the Weil-Petersson geometry of the moduli space of
polarized CY manifolds can be found in \cite{Lu} and \cite{LS}. In those two
papers important results are obtained.

\subsection{Global Moduli}

\begin{definition}
\label{Teich}We will define the Teichm\"{u}ller space $\mathcal{T}$(M) of a CY
manifold M as follows: $\mathcal{T}($M$):=\mathcal{I}($M$)/Diff_{0}($M$),$
\textit{where}\
\[
\mathcal{I}(\text{M}):=\left\{  \text{all integrable complex structures on
M}\right\}
\]
\textit{and } Diff$_{0}$(M) \textit{is the group of diffeomorphisms isotopic
to identity. The action of the group }$Diff_{0}$\textit{(M}$)$ \textit{is
defined as follows; Let }$\phi\in$Diff$_{0}$(M) \textit{then }$\phi$
\textit{acts on integrable complex structures on M by pull back, i.e. if }
$I\in C^{\infty}($M$,Hom(T($M$),T($M$)),$ \textit{then we define }
$\phi(I_{\tau})=\phi^{\ast}(I_{\tau}).$
\end{definition}

We will call a pair $($M$;\gamma_{1},...,\gamma_{b_{n}})$ a marked CY manifold
where M is a CY manifold and $\{\gamma_{1},...,\gamma_{b_{n}}\}$ is a basis of
$H_{n}$(M,$\mathbb{Z}$)/Tor.

\begin{remark}
\label{mark}Let $\mathcal{K}$ be the Kuranishi space. It is easy to see that
if we choose a basis of $H_{n}$(M,$\mathbb{Z}$)/Tor in one of the fibres of
the Kuranishi family $\pi:\mathcal{X}_{\mathcal{K}}\mathcal{\rightarrow K}$
then all the fibres will be marked, since as a $C^{\infty}$ manifold
$\mathcal{X}_{\mathcal{K}}\approxeq$M$\times\mathcal{K}$.
\end{remark}

In \cite{LTYZ} the following Theorem was proved:

\begin{theorem}
\label{teich}There exists a family of marked polarized CY manifolds
\begin{equation}
\mathcal{Z}_{L}\mathcal{\rightarrow}\mathfrak{T}(\text{M}), \label{fam2}%
\end{equation}
which possesses the following properties: \textbf{a)} It is effectively
parametrized, \textbf{b) }For any marked CY manifold M of fixed topological
type for which the polarization class $L$ defines an imbedding into a
projective space $\mathbb{CP}^{N},$ there exists an isomorphism of it (as a
marked CY manifold) with a fibre M$_{s}$ of the family $\mathcal{Z}_{L}.$
\textbf{c) }The base has dimension $h^{n-1,1}.$
\end{theorem}

\begin{corollary}
\label{teich1}Let $\mathcal{Y\rightarrow}\mathfrak{X}$ be any family of marked
polarized CY manifolds, then there exists a unique holomorphic map
$\phi:\mathfrak{X}\rightarrow\mathfrak{T}($M$)$ up to a biholomorphic map
$\psi$ of M which induces the identity map on $H_{n}($M$,\mathbb{Z}).$
\end{corollary}

From now on we will denote by $\mathcal{T}$(M) the irreducible component of
the Teichm\"{u}ller space that contains our fixed CY manifold M.

\begin{definition}
We will define the mapping class group $\Gamma_{1}($M$)$ of any compact
C$^{\infty}$ manifold M as follows:
\[
\Gamma_{1}(\text{M})=Diff_{+}\left(  \text{M}\right)  /Diff_{0}\left(
\text{M}\right)  ,
\]
where $Diff_{+}($M$)$ is the group of diffeomorphisms of M preserving the
orientation of M and $Diff_{0}($M$)$ is the group of diffeomorphisms isotopic
to identity.
\end{definition}

\begin{definition}
Let $L\in H^{2}($M$,\mathbb{Z})$ be the imaginary part of a K\"{a}hler metric.
We will denote by
\[
\Gamma_{2}:=\{\phi\in\Gamma_{1}(M)|\phi(L)=L\}.
\]

\end{definition}

It is a well know fact that the moduli space of polarized algebraic manifolds
$\mathcal{M}_{L}($M$)=\mathcal{T}($M$)/\Gamma_{2}.$ In \cite{LTYZ} the
following fact was established:

\begin{theorem}
\label{Vie}There exists a subgroup of finite index $\Gamma_{L}$ of
$\ \Gamma_{2}$ such that $\Gamma_{L}$ acts freely on $\mathcal{T}$(M) and
$\Gamma\backslash\mathcal{T}($M$)=\mathfrak{M}_{L}($M$)$ is a non-singular
quasi-projective variety. Over $\mathfrak{M}_{L}$(M) there exists a family of
polarized CY manifolds $\pi:\mathcal{M}\rightarrow\mathfrak{M}_{L}($M$).$
\end{theorem}

\begin{remark}
\label{Vie1}Theorem \ref{Vie} implies that we constructed a family of
non-singular CY manifolds
\begin{equation}
\pi:\mathcal{X\rightarrow}\mathfrak{M}_{L}(\text{M}) \label{FAM}%
\end{equation}
over a quasi-projective non-singular variety $\mathfrak{M}_{L}($M$)$. Moreover
it is easy to see that $\mathcal{X\subset}\mathbb{CP}^{N}\times\mathfrak{M}%
_{L}($M$).$ \textit{So} $\mathcal{X}$ \ \textit{is also quasi-projective. From
now on we will work only with this family.}
\end{remark}

\begin{remark}
\label{Vie2}Theorem \ref{Vie} implies that $\mathfrak{M}_{L}($M$)$ is a
quasi-projective non-singular variety. Using Hironaka's resolution theorem, we
can find a compactification $\overline{\mathfrak{M}_{L}(\text{M})}$ of
$\mathfrak{M}_{L}($M$)$ such that $\overline{\mathfrak{M}_{L}(\text{M}%
)}-\mathfrak{M}_{L}($M$)=\mathfrak{D}$ is a divisor with normal crossings. We
will call $\mathfrak{D}$ the discriminant divisor.
\end{remark}

\subsection{Affine Flat Coordinates around Points at Infinity}

\begin{theorem}
\label{FCS}Let $U_{\infty}=D^{N}\subset$ $\overline{\mathfrak{M}_{L}%
(\text{M})}$ be some open polydisk containing the point $\tau_{\infty}%
\in\mathfrak{D}.$ Suppose that
\[
U_{\infty}-\left(  U_{\infty}\cap\mathfrak{D}\right)  =\left(  D^{\ast
}\right)  ^{k}\times D^{N-k}.
\]
According to the results proved in \cite{LTYZ} there exists a complete family
of polarized CY manifolds
\begin{equation}
\pi:\mathcal{X}\rightarrow U_{\infty}-\left(  U_{\infty}\cap\mathfrak{D}%
\right)  =\left(  D^{\ast}\right)  ^{k}\times D^{N-k}. \label{ifam}%
\end{equation}
Let $\omega_{\mathcal{X}/U_{\infty}}$ be the relative dualizing line bundle.
Then there exists a coordinate system $(\tau^{1},...,\tau^{k},t^{1}%
,...,t^{N-k})$ on the universal cover $\left(  \mathrm{U}\right)  ^{k}\times
D^{N-k}$ of $\left(  D^{\ast}\right)  ^{k}\times D^{N-k}$ and a global section
$\omega_{\tau}\in\Gamma\left(  \left(  \mathrm{U}\right)  ^{k},\pi_{\ast
}\omega_{\mathcal{X}/\left(  \mathrm{U}\right)  ^{k}}\right)  $ such that$:$%
\[
\omega_{\tau}=\omega_{0}+%
{\displaystyle\sum\limits_{i=1}^{k}}
\omega_{i,0}(n-1,1)\tau^{i}+%
{\displaystyle\sum\limits_{i\leq j=1}^{k}}
\omega_{ij,0}(n-2,2)\tau^{i}\tau^{j}+O(3)+
\]%
\begin{equation}%
{\displaystyle\sum\limits_{j=1}^{n\_k}}
\omega_{j,0}(n-1,1)t^{j}+%
{\displaystyle\sum\limits_{i\leq j=1}^{N-k}}
\omega_{ij,0}(n-2,2)t^{i}t^{j}+O(3). \label{LCS0}%
\end{equation}

\end{theorem}

\textbf{Proof: }The proof of Theorem \ref{FCS} is based on the results
obtained in \cite{LTY}.

\begin{lemma}
\label{fcs}Suppose that $\tau_{\infty}=0\subset U_{\infty}$ and $D$ is an open
disk in $U_{\infty}$ containing $0$. Suppose that the monodromy operator $T$
of the restriction of the family $\left(  \ref{ifam}\right)  $ on
$D-D\cap\mathfrak{D}$ is of infinite order. Then there exists a non zero
section
\[
\omega_{\tau}\in\Gamma\left(  U_{\infty}-\left(  U_{\infty}\cap\mathfrak{D}%
\right)  ,\pi_{\ast}\omega_{\mathcal{X}\left/  U_{\infty}-\left(  U_{\infty
}\cap\mathfrak{D}\right)  \right.  }\right)  ,
\]
such that%
\begin{equation}%
{\displaystyle\int\limits_{\gamma_{0}}}
\omega_{\tau}=1, \label{FCY0}%
\end{equation}
where $\gamma_{0}$ a a primitive invariant vanishing cycle with respect to the
monodromy operator $T.$
\end{lemma}

\textbf{Proof: }Let us consider the family $\left(  \ref{ifam}\right)  .$
Since we assumed that $D\cap U_{\infty}$ is any open disk and that $U_{\infty
}$ is a polydisk, then we can construct a non zero family of holomorphic forms
$\Omega_{t}$ over $D\cap U_{\infty}$ according to \cite{LTY}. So we can
analytically extend this family to a family of holomorphic forms $\Omega
_{\tau}$ over $U_{\infty}.$ Thus we get:
\[
\Omega_{\tau}\in\Gamma\left(  \pi^{-1}\left(  U_{\infty}-D\cap U_{\infty
}\right)  ,\omega_{\mathcal{X}/U_{\infty}}\right)
\]
such that at each $\tau\in U_{\infty}-D\cap U_{\infty},$ $\Omega_{\tau}\neq0.$
According to Theorem \textbf{37} proved in \cite{LTY} we have
\begin{equation}%
{\displaystyle\int\limits_{\gamma_{0}}}
\Omega_{t}\neq0\text{ and }\underset{t\rightarrow0}{\lim}%
{\displaystyle\int\limits_{\gamma_{0}}}
\Omega_{t}\neq0 \label{FCY1}%
\end{equation}
for $t\in D.$ $\left(  \ref{FCY1}\right)  $ implies that the function
$\phi(t)=%
{\displaystyle\int\limits_{\gamma_{0}}}
\Omega_{\tau}$ is different from zero on $U_{\infty}-D\cap U_{\infty}.$ Then
we can define $\omega_{\tau}=\frac{\Omega_{\tau}}{\phi(\tau)}.$ Clearly the
family of holomorphic $n$-forms $\omega_{\tau}$ satisfies $\left(
\ref{FCY0}\right)  .$ Lemma \ref{fcs} is proved. $\blacksquare$

\begin{lemma}
\label{fcsa}Suppose that the monodromy operator $T$ of the restriction of the
family $\left(  \ref{ifam}\right)  $ on $D-D\cap\mathfrak{D}=\tau_{\infty}$ is
of finite order $m.$ Then there exists a $n-$cycle $\gamma_{0}$ and a non zero
section $\omega_{\tau}\in\Gamma\left(  U_{\infty},\pi_{\ast}\omega
_{\mathcal{X}/U_{\infty}}\left\langle \log\mathfrak{D}\right\rangle \right)
,$ such that on $U_{\infty}$ we have
\begin{equation}
\underset{\tau\rightarrow0}{\lim}%
{\displaystyle\int\limits_{\gamma_{0}}}
\omega_{\tau}=1. \label{one}%
\end{equation}

\end{lemma}

\textbf{Proof: }Let $\phi_{m}:D\rightarrow D$ be the map $t\rightarrow t^{m}.$
Let us pullback the restriction of the family $\left(  \ref{ifam}\right)  $ by
$\phi_{m}.$ Then the monodromy operator $T$ of the new family will be the
identity. Then we can choose a $n-$cycle $\gamma_{0}$ such that%
\[
\underset{t\rightarrow0}{\lim}%
{\displaystyle\int\limits_{\gamma_{0}}}
\Omega_{t}\neq0.
\]
The family of holomorphic forms $\Omega_{t}$ can be prolong to a family
$\Omega_{\tau}$ over $U_{\infty}-\left(  U_{\infty}\cap\mathfrak{D}\right)  $
such that $\phi(\tau):=%
{\displaystyle\int\limits_{\gamma_{0}}}
\Omega_{\tau}$ is a non zero function on $U_{\infty}.$ Then $\omega_{\tau
}:=\frac{\Omega_{\tau}}{\phi(\tau)}$ satisfies
\[
\underset{\tau\rightarrow0}{\lim}%
{\displaystyle\int\limits_{\gamma_{0}}}
\omega_{\tau}=1
\]
Lemma \ref{fcsa} is proved. $\blacksquare$

We define the flat affine coordinates
\[
\left(  \tau^{1},...,\tau^{k},t^{1},...,t^{N-k}\right)
\]
in $\left(  \mathrm{U}\right)  ^{k}\times D^{N-k}$ as follows: Let
$\omega_{\tau}$ be the family of holomorphic $n-$ forms defined on $U_{\infty
}$ by Lemmas \ref{fcs} and \ref{fcsa}. Local Torelli Theorem implies that we
can choose a basis of cycles
\[
\left(  \gamma_{0},\gamma_{1},....,\gamma_{N},\gamma_{N+1},...,\gamma
_{2N+1},...\gamma_{b_{n}}\right)
\]
of $H_{n}\left(  \text{M,}\mathbb{Z}\right)  $ satisfying
\[
\left\langle \gamma_{i},\gamma_{j}\right\rangle _{\text{M}}=0,\text{
}\left\langle \gamma_{i},\gamma_{2N+1-j}\right\rangle _{\text{M}}=\delta_{ij}%
\]
for $i=0,...,k;$ $j=1,...,N-k$ such that if
\begin{equation}
\tau^{i}:=%
{\displaystyle\int\limits_{\gamma_{i}}}
\omega_{\tau},\text{ }i=1,...,k\text{ and }t^{j}:=%
{\displaystyle\int\limits_{\gamma_{j+k}}}
\omega_{\tau},\text{ }j=1,...,N-k \label{lcs}%
\end{equation}
then $(\tau^{1},...,\tau^{k},t^{1},...,t^{N-k})$ will be a local coordinate
system in $\left(  \mathrm{U}\right)  ^{k}\times\left(  D\right)  ^{N-k}.$

\begin{lemma}
\label{fcs0}Let $0\in(\mathrm{U)}^{k}$ be any fixed point. Then the Taylor
expansion of the family of holomorphic $n$ forms $\omega_{\tau}$ constructed
in Lemmas \ref{fcs} and \ref{fcsa} satisfies $\left(  \ref{LCS0}\right)  .$
\end{lemma}

\textbf{Proof: }We know from \cite{To89} that we can identify the tangent
space at $0\in U_{\infty}-\left(  U_{\infty}\cap\mathfrak{D}\right)  $ with
$H^{1}($M$_{0},T_{\text{M}_{0}}^{1,0}).$ The contraction with $\omega_{0}$
defines an isomorphism%
\[
H^{1}\left(  \text{M}_{0},T_{\text{M}_{0}}^{1,0}\right)  \approxeq
H^{1}\left(  \text{M}_{0},\Omega_{\text{M}_{0}}^{n-1,0}\right)  .
\]
Thus the tangent vectors
\[
\phi_{i}=\frac{\partial}{\partial\tau^{i}}\in T_{0,U_{\infty}}=H^{1}\left(
\text{M}_{0},\Omega_{\text{M}_{0}}^{n-1,0}\right)
\]
can be identified with classes of cohomologies $\omega_{i,0}(n-1,1):=\omega
_{0}\lrcorner\phi_{i}$ of type $(n-1,1).$ Griffiths' transversality implies
that for $t=0$ we have%
\begin{equation}
\left(  \frac{\partial}{\partial\tau^{i}}\omega_{\tau}\right)  |_{\tau
=0}=a_{0}\omega_{0}+\omega_{0}\lrcorner\phi_{i}=a_{0}\omega_{0}+\omega
_{i,0}(n-1,1) \label{LCS1a}%
\end{equation}
and%
\[
\left(  \frac{\partial^{2}}{\partial\tau^{i}\partial\tau^{j}}\omega_{\tau
}\right)  |_{\tau=0}=
\]%
\begin{equation}
\text{ }a_{i,j}(0)\omega_{0}+b_{i,j}(0)\left(  \omega_{i,j,0}(n-1,1)\right)
+c_{ij}(0)\omega_{i,j,0}(n-2,2). \label{LCS1}%
\end{equation}

\begin{proposition}
\label{fcso1}We have $a_{0}=a_{i,j}(0)=b_{i,j}(0)=0$ and $c_{ij}%
(0)=const\neq0$ in the expression $\left(  \ref{LCS1}\right)  .$
\end{proposition}

\textbf{Proof: }The definition of the coordinates $(\tau^{1},...,\tau^{k})$
and$\ \left(  \ref{LCS1a}\right)  $ and $\left(  \ref{one}\right)  $ imply
that%
\begin{equation}
a_{0}=0.\label{LCSa}%
\end{equation}
From $\left(  \ref{LCS1}\right)  $ and $\left(  \ref{LCSa}\right)  $ we can
conclude that for $1\leq i\leq k$ and $t=0$ we have%
\[
\omega_{\tau}\left\vert _{\left(  \mathrm{U}\right)  ^{k}}\right.  =\omega
_{0}+%
{\displaystyle\sum\limits_{i=1}^{k}}
\tau^{i}\left(  \omega_{0}\lrcorner\phi_{i}\right)  +
\]%
\begin{equation}
\frac{1}{2}%
{\displaystyle\sum\limits_{i,j=1}^{k}}
\left(  b_{ij}(0)\omega_{i,j,0}(n-1,1)+c_{ij}(0)\left(  \omega_{0}%
\lrcorner\phi_{i}\lrcorner\phi_{j}\right)  \right)  \tau^{i}\tau
^{j}+...,\label{LCSb}%
\end{equation}
where $b_{ij}(0)$ and $c_{ij}(0)$ are constants$.$ So $\left(  \ref{one}%
\right)  $ implies%
\[%
{\displaystyle\int\limits_{\gamma_{0}}}
\frac{\partial}{\partial\tau^{i}}\omega_{\tau}=%
{\displaystyle\int\limits_{\gamma_{0}}}
\frac{\partial^{2}}{\partial\tau^{i}\partial\tau^{j}}\omega_{\tau}=0.
\]
Since
\[
\left(  \frac{\partial}{\partial\tau^{i}}\omega_{\tau}\right)  \left\vert
_{\tau=0}\right.  =\omega_{0}\lrcorner\phi_{i}:=\omega_{i,0}(n-1,1),
\]
and
\[
\omega_{\tau}\left\vert _{\left(  \mathrm{U}\right)  ^{k}}\right.  =\omega
_{0}+%
{\displaystyle\sum\limits_{i=1}^{k}}
a_{i}\tau^{i}\left(  \omega_{i,0}(n-1,1)\right)  +...
\]
we deduce that
\begin{equation}%
{\displaystyle\int\limits_{\gamma_{0}}}
[\omega_{i,0}(n-1,1)]=0\text{ and }%
{\displaystyle\int\limits_{\gamma_{j}}}
[\omega_{i,0}(n-1,1)]=\delta_{ij}.\label{LCS}%
\end{equation}
Thus $\left(  \ref{LCS}\right)  $ and $\left(  \ref{LCSb}\right)  $ imply that
$a_{i}=1.$

The relations $\left(  \ref{lcs}\right)  $ and $\left(  \ref{LCSb}\right)  $
imply that%
\begin{equation}%
{\displaystyle\int\limits_{\gamma_{k}}}
\omega_{i,j,0}(n-1,1)=0 \label{lcs1}%
\end{equation}
for any $\gamma_{k}$ such that $%
{\displaystyle\int\limits_{\gamma_{k}}}
\omega_{\tau}=\tau^{k}.$Indeed $\left(  \ref{LCS}\right)  $ implies that for
any non zero closed form $\omega(n-1,1)$ of type $(n-1,1)$ there exists
$\gamma_{k}$ such that
\begin{equation}%
{\displaystyle\int\limits_{\gamma_{k}}}
\omega(n-1,1)\neq0. \label{lcs2}%
\end{equation}
Thus $\left(  \ref{lcs1}\right)  $ and $\left(  \ref{lcs2}\right)  $ imply
that $b_{i,j}(0)=0.$ Proposition \ref{fcso1} is proved. $\blacksquare$

Proposition \ref{fcso1} implies Lemma \ref{fcs0}. $\blacksquare$ Lemma
\ref{fcs0} implies Theorem \ref{FCS}. $\blacksquare$

\begin{remark}
Theorem \ref{FCS} states that the coordinates used in the special geometry and
the flat affine coordinates introduced in \cite{To89} by Theorems \ref{tod1}
and \ref{forms} are the same. This fact is mentioned in \textbf{\#5.1 }of
\cite{BCOV}. In the same paper the authors referred to [30] (private
communication by Tian) for the introduction of the flat affine coordinates.
\end{remark}

\section{Metrics on Vector Bundles with Logarithmic Growth}

\subsection{Mumford Theory}

In this \textbf{Section }we are going to recall some definitions and results
from \cite{Mu}. Let X be a quasi-projective variety. Let $\overline{\text{X}}$
be a projective compactification of X such that $\overline{\text{X}}%
-$X$=\mathfrak{D}_{\infty}$ is a divisor with normal crossings. The existence
of such compactification follows from the Hironaka's results. We will look at
polydisk $D^{N}\subset\overline{\text{X}},$ where D is the unit disk,
$N=\dim\overline{\text{X}}$ such that
\[
D^{N}\cap\text{X}=(D^{\ast})^{k}\times\text{D}^{N-k},
\]
where $D^{\ast}=D-0$ and $q$ is the coordinate in $D.$ On $D^{\ast}$ we have
the Poincare metric
\[
ds^{2}=\frac{\left\vert dq\right\vert ^{2}}{\left\vert q\right\vert
^{2}\left(  \log\left\vert q\right\vert \right)  ^{2}}.
\]
On the unit disk D we have the simple metric $\left\vert dt\right\vert ^{2}.$
The product metric on ($D^{\ast}$)$^{k}\times$D$^{N-k}$ we will call
$\omega^{(P)}.$

A complex-valued C$^{\infty}$ p-form $\eta$ on X is said to have Poincare
growth on $\overline{\text{X}}-$X if there is a set of if for a covering
$\left\{  \mathcal{U}_{\alpha}\right\}  $ by polydisks of $\overline{\text{X}%
}-$X such that in each \textit{\ }$\mathcal{U}_{\alpha}$ the following
estimate holds:
\begin{equation}
\left\vert \eta\left(  q^{1},...,q^{k},t^{k+1},...,t^{N}\right)  \right\vert
\leq C_{\alpha}\left\Vert \omega_{\mathcal{U}_{\alpha}}^{(p)}(q^{1}%
,\overline{q^{1}})\right\Vert ^{2}...\left\Vert \omega_{\mathcal{U}_{\alpha}%
}^{(p)}(q^{k},\overline{q^{k}})\right\Vert ^{2} \label{PC}%
\end{equation}
where
\[
\left\Vert \omega_{\mathcal{U}_{\alpha}}^{(p)}(q^{i},\overline{q^{i}%
})\right\Vert ^{2}=\frac{|q^{i}|^{2}}{\left\vert q^{i}\right\vert ^{2}\left(
\log\left\vert q^{i}\right\vert \right)  ^{2}}%
\]
This property is independent of the covering $\left\{  \mathcal{U}_{\alpha
}\right\}  $ of $\ $X but depends on the compactification $\overline{\text{X}%
}.$ If $\eta_{1}$ and $\eta_{2}$ both have Poincare growth on $\overline
{\text{X}}-$X then so does $\eta_{1}\wedge\eta_{2}$. A complex valued
C$^{\infty}$ p-form\textit{\ }$\eta$ on\textit{\ }$\overline{\text{X}}%
$\textit{\ }will be called "good" on\textit{ }$\overline{\text{X}}%
$\textit{\ if both }$\eta$\textit{\ }and $d\eta$ have Poincare growth\textit{.
}

An important property of Poincare growth is the following:

\begin{theorem}
\label{Mum}Suppose that the $\eta$ is a p-form with a Poincare growth on
$\overline{\text{X}}-$X$=\mathfrak{D}_{\infty}\mathfrak{.}$ Then \textit{for
every C}$^{\infty}$ $(r-p)$ \textit{form} $\psi$ \textit{on \ }$\overline
{\text{X}}$ \textit{we have:}
\[
\int_{\overline{\text{X}}}\left\vert \eta\wedge\psi\right\vert <\infty.
\]
\textit{Hence, }$\eta$ \textit{defines a current }$\left[  \eta\right]  $
\textit{on }$\overline{\text{X}}.$
\end{theorem}

\textbf{Proof:}For the proof see \cite{Mu}. $\blacksquare$

\begin{definition}
\label{Gforms}Let $\mathcal{E}$ be a vector bundle on X with a Hermitian
metric h. We will call h a good metric on $\overline{\text{X}}$ if the
following holds. \textbf{1. }\textit{If for all x}$\in\overline{\text{X}}-X,$
\textit{there exist sections }%
\[
e_{1},...,e_{m}\in\mathcal{E}\left\vert _{D^{N}-(D^{N}\cap\mathfrak{D}%
_{\infty})}\right.
\]
\textit{of} $\mathcal{E}$ \ \textit{which form a basis of }$\mathcal{E}%
\left\vert _{D^{N}-(D^{N}\cap\mathfrak{D}_{\infty})}\right.  .$ \textbf{2.
}\textit{In a neighborhood }$D^{N}$ \textit{of x}$\in$ $\overline{\text{X}}-$X
$\ $\textit{in which }%
\[
D^{N}\cap\text{X}=(D^{\ast})^{k}\times D^{N-k}%
\]
\textit{and} $\overline{\text{X}}-$X$=\mathfrak{D}_{\infty}$ \textit{is given
by }%
\[
t^{1}\times...\times t^{N}=0\mathit{\ }\
\]
\textit{the metric }h$_{i\overline{j}}=$h($e_{i},e_{j}$)\textit{\ has the
following properties: }\textbf{a. }
\begin{equation}
\left\vert h_{i\overline{j}}\right\vert \leq C\left(  \sum_{i=1}^{k}%
\log\left\vert q^{i}\right\vert \right)  ^{2m},\text{ }\left(  \det\left(
h\right)  \right)  ^{-1}\leq C\left(  \sum_{i=1}^{k}\log\left\vert
q^{i}\right\vert \right)  ^{2m}\label{GM}%
\end{equation}
\textit{for some }$C>0$ and $m\geq0.$ \textbf{b.}\ \textit{The 1-forms
}$\left(  \left(  dh\right)  h^{-1}\right)  $\textit{\ are good forms
on}$\overline{\text{X}}\cap D^{N}.$
\end{definition}

It is easy to prove that there exists a unique extension $\overline
{\mathcal{E}}$ of $\mathcal{E}$ \ on $\overline{\text{X}},$ i.e.
$\overline{\mathcal{E}}$ is defined locally as\ holomorphic sections of
$\mathcal{E}$ which have a finite norm in h.

\begin{theorem}
\label{Mum100}Let $\left(  \mathcal{E},h\right)  $ be a vector bundle with a
good metric on $\overline{\text{X}}$, then the Chern classes $c_{k}%
$($\mathcal{E}$,h) are good forms on $\overline{\text{X}}$ \textit{and the
currents }$\left[  c_{k}\left(  \mathcal{E},h\right)  \right]  $
\textit{represent the cohomology classes }$c_{k}(\mathcal{E},$h$)\in
H^{2k}\left(  \overline{\text{X}},\mathbb{Z}\right)  .$
\end{theorem}

\textbf{Proof: }For the proof see \cite{Mu}. $\blacksquare$

\subsection{Example of a Good Metric}

\begin{theorem}
\label{gmet}Let $\pi:\mathrm{U}_{1}\times...\times\mathrm{U}_{k}%
\rightarrow\left(  D^{\ast}\right)  ^{k}$ be the uniformization map, where
\textrm{U} is the unit disk. Suppose that
\[
\kappa_{\infty}\in\left(  \partial\left(  \mathrm{U}\right)  \right)
^{k}=\left(  S^{1}\right)  ^{k},
\]
Let $h$ be a metric on the line bundle $\mathcal{L}\rightarrow\left(  D^{\ast
}\right)  ^{k}.$ Let
\[
\left\{  \tau_{m}\right\}  \in\mathrm{U}_{1}\times...\times\mathrm{U}_{k}%
\]
be any sequence such that%
\[
\underset{m\rightarrow\infty}{\lim}\tau_{m}=\kappa_{\infty}\in\overline
{\mathrm{U}_{1}}\times...\times\overline{\mathrm{U}_{k}}-\mathrm{U}_{1}%
\times...\times\mathrm{U}_{k}%
\]
and%
\[
\underset{m\rightarrow\infty}{\lim}\pi(\tau_{m})=0\in\left(  D\right)
^{k}=\overline{\left(  D^{\ast}\right)  ^{k}}.
\]
Suppose that $\pi^{\ast}(h)=$ $h_{\mathrm{U}^{k}}$ is defined on
\textrm{U}$_{1}\times...\times\mathrm{U}_{k}$ as follows:
\begin{equation}
h_{\mathrm{U}^{k}}:=%
{\displaystyle\sum\limits_{i=1}^{k}}
\left(  1-|\tau^{i}|^{2}\right)  +\phi(\tau,\overline{\tau}), \label{gmet0}%
\end{equation}
where $\phi(\tau,\overline{\tau})$ is a bounded $C^{\infty}$ function on
$\left(  \mathrm{U}\right)  ^{k}$ and%
\[
\underset{m\rightarrow\infty}{\lim}\phi(\tau_{m},\overline{\tau_{m}%
})=\underset{m\rightarrow\infty}{\lim}h_{\mathrm{U}^{k}}(\tau_{m}%
,\overline{\tau_{m}})=0.
\]
Then $h$ is a good metric in the sense of Mumford on the line bundle
$\mathcal{L}\rightarrow\left(  D^{\ast}\right)  ^{k}.$
\end{theorem}

\textbf{Proof: }We need to show that $h$ satisfies the conditions $\left(
\ref{GM}\right)  $ and that $\partial\log h_{\mathrm{U}^{k}}$ is a good form.
The conditions $\left(  \ref{GM}\right)  $ followed immediately from the
expression $\left(  \ref{gmet}\right)  $ for the metric defined by
$h_{\mathrm{U}^{k}}.$ We need to show that $h$ satisfies $\left(
\ref{PC}\right)  ,$ i.e. $\partial\log h$ is a good form.

\begin{lemma}
\label{gmetb}The $(1,0)$ form $\partial\log h$ is a good form.
\end{lemma}

\textbf{Proof: } The definition of a good form implies that $\partial\log h$
is a good form on $\left(  D^{\ast}\right)  ^{k}$ if and only iff
$\partial\log h_{\left(  \mathrm{U}\right)  ^{k}}$ satisfies on the universal
cover $\left(  \mathrm{U}\right)  ^{k}$ of $(D^{\ast})^{k}$ the following
inequalities on each unit disk $\mathrm{U}_{i}\subset\left(  \mathrm{U}%
\right)  ^{k}$:
\begin{equation}
0\leq\frac{\frac{\partial}{\partial\tau^{i}}h_{\mathrm{U}_{i}}}{h_{\mathrm{U}%
_{i}}}\overline{\frac{\frac{\partial}{\partial\tau^{i}}h_{\mathrm{U}_{i}}%
}{h_{\mathrm{U}_{i}}}}\leq c\frac{1}{\left(  1-|\tau^{i}|^{2}\right)  ^{2}%
}\label{gmet1a}%
\end{equation}
and%
\begin{equation}
0\leq\overline{\frac{\partial}{\partial\tau^{i}}}\left(  \frac{\frac{\partial
}{\partial\tau^{i}}h_{\mathrm{U}_{i}}}{h_{\mathrm{U}_{i}}}\right)  \leq c%
{\displaystyle\sum\limits_{i=1}^{k}}
\frac{1}{\left(  1-|\tau^{i}|^{2}\right)  ^{2}},\label{gmet1}%
\end{equation}
where $c>0.$ This statement follows from the fact that the pullback of the
metric with a constant curvature on $\left(  D^{\ast}\right)  ^{k}$ is the
Poincare metric on $\left(  \mathrm{U}\right)  ^{k},$ i.e.
\[
\pi^{\ast}\left(
{\displaystyle\sum\limits_{i=1}^{k}}
\frac{\left\vert dq^{i}\right\vert ^{2}}{\left\vert q^{i}\right\vert
^{2}\left(  \log\left\vert q^{i}\right\vert \right)  ^{2}}\right)  =%
{\displaystyle\sum\limits_{i=1}^{k}}
\frac{\left\vert d\tau^{i}\right\vert ^{2}}{\left(  1-|\tau^{i}|^{2}\right)
^{2}}%
\]
and $\partial\log\pi^{\ast}(h)=\partial\log h_{\left(  \mathrm{U}\right)
^{k}}$.

\begin{proposition}
\label{gmetb1}The form $\partial\log h_{\left(  \mathrm{U}\right)  ^{k}}$
satisfies $\left(  \ref{gmet1a}\right)  $ and $\left(  \ref{gmet1}\right)  .$
\end{proposition}

\textbf{Proof: }$\left(  \ref{gmet1a}\right)  $ and $\left(  \ref{gmet1}%
\right)  $ will follow if we prove that the restriction of $\partial\log
h_{\left(  \mathrm{U}\right)  ^{k}}$ on each $\mathrm{U}_{i}$ satisfies
$\left(  \ref{gmet1a}\right)  $ and $\left(  \ref{gmet1}\right)  .$ Direct
computations show that the expression $\left(  \ref{gmet0}\right)  $ of
$h_{\left(  \mathrm{U}\right)  ^{k}}$ implies that we have :%
\begin{equation}
h_{i}\left(  \tau^{i},\overline{\tau^{i}}\right)  :=h_{\left(  \mathrm{U}%
\right)  ^{k}}|_{\mathrm{U}_{i}}=\left(  1-|\tau|^{2}\right)  +\phi_{i}\left(
\tau^{i},\overline{\tau^{i}}\right)  >0,\label{gmet2}%
\end{equation}
where%
\[
\underset{\tau^{i}\rightarrow\kappa_{\infty}^{i}}{\lim}\phi_{i}(\tau
^{i},\overline{\tau^{i}})=\underset{\tau^{i}\rightarrow\kappa_{\infty}^{i}%
}{\lim}h_{i}(\tau^{i},\overline{\tau^{i}})=0
\]
and $\phi_{i}(\tau^{i},\overline{\tau^{i}})$ is a bounded $C^{\infty}$
function on \textrm{U}$_{i}.$ $\left(  \ref{gmet2}\right)  $ implies:%
\begin{equation}
0\leq\left(  1-|\tau^{i}|^{2}\right)  \leq C_{i}h_{i}\left(  \tau
^{i},\overline{\tau^{i}}\right)  ,\label{gmet3}%
\end{equation}
where $C_{i}>0.$ Thus we get from $\left(  \ref{gmet3}\right)  $ that if
$\tau_{0}$ is any complex number such that $\left\vert \tau_{0}\right\vert =1$
then the limit%
\[
\underset{\tau^{i}\rightarrow\tau_{0}}{\lim}\frac{\left(  1-|\tau^{i}%
|^{2}\right)  }{h_{i}\left(  \tau^{i},\overline{\tau^{i}}\right)  }%
\]
exists and
\begin{equation}
0\leq\underset{\tau^{i}\rightarrow\tau_{0}^{i}}{\lim}\frac{\left(  1-|\tau
^{i}|^{2}\right)  }{h_{i}\left(  \tau^{i},\overline{\tau^{i}}\right)
}=c.\label{gmet4}%
\end{equation}
Direct computations show that
\[
\frac{\partial}{\partial\tau^{i}}\log h_{h_{\mathrm{U}_{i}}}=\frac
{\frac{\partial}{\partial\tau^{i}}\left(  1-|\tau^{i}|^{2}+\phi_{i}\left(
\tau^{i},\overline{\tau^{i}}\right)  \right)  }{\left(  1-|\tau^{i}|^{2}%
+\phi_{i}\left(  \tau^{i},\overline{\tau^{i}}\right)  \right)  }=
\]%
\begin{equation}
\frac{\left(  -\overline{\tau^{i}}+\frac{\partial}{\partial\tau^{i}}\phi
_{i}\left(  \tau^{i},\overline{\tau^{i}}\right)  \right)  }{\left(
1-|\tau^{i}|^{2}+\phi_{i}\left(  \tau^{i},\overline{\tau^{i}}\right)  \right)
}.\label{gmet4a}%
\end{equation}
We derive from $\left(  \ref{gmet4}\right)  ,$ $\left(  \ref{gmet4a}\right)  $
and the fact that $\phi_{i}$ is bounded $C^{\infty}$ function on
\textrm{U}$_{i}$ that we have:
\[
0\leq\frac{\frac{\partial}{\partial\tau^{i}}h_{h_{\mathrm{U}_{i}}}%
}{h_{\mathrm{U}_{i}}}\overline{\frac{\frac{\partial}{\partial\tau^{i}%
}h_{\mathrm{U}_{i}}}{h_{\mathrm{U}_{i}}}_{_{i}}}\leq c_{1}\frac{1}{\left(
1-|\tau^{i}|^{2}\right)  ^{2}}%
\]
and%
\begin{equation}
0\leq\left\vert \overline{\frac{\partial}{\partial\tau^{i}}}\left(
\frac{\partial h_{\mathrm{U}_{i}}}{h_{\mathrm{U}_{i}}}\right)  \right\vert
\leq c_{1}\frac{1}{\left(  1-|\tau^{i}|^{2}\right)  ^{2}},\label{gmet5}%
\end{equation}
where $c_{1}>0.$ Thus $\left(  \ref{gmet5}\right)  $ implies that
$\partial\log h$ defines a good form on the line bundle $\mathcal{L}$
restricted on $(D^{\ast})^{k}$. Lemma \ref{gmetb} is proved. $\blacksquare$

Lemma \ref{gmetb} implies Theorem \ref{gmet}. $\blacksquare$

\section{Applications of Mumford Theory to the Moduli of CY}

\subsection{The $L^{2}$ Metric is Good}

We are going to prove the following result:

\begin{theorem}
\label{Nik}The natural $L^{2}$ metric :
\begin{equation}
h(\tau,\overline{\tau})=\left\Vert \omega_{\tau}\right\Vert ^{2}%
:=(-1)^{\frac{n(n-1)}{2}}\left(  \sqrt{-1}\right)  ^{n}\int_{\text{M}}%
\omega_{\tau}\wedge\overline{\omega_{\tau}} \label{W-P}%
\end{equation}
on $\pi_{\ast}\left(  \omega_{\mathcal{X}\text{(M)/}\mathfrak{M}_{L}%
(\text{M})}\right)  \rightarrow\mathfrak{M}_{L}(M)$ \textit{is a good metric.}
\end{theorem}

\textbf{Outline of the proof of Theorem }\ref{Nik}. Let\textbf{ }$\left(
D\right)  ^{N}$ be a polydisk in $\overline{\mathfrak{M}_{L}\left(
\text{M}\right)  }$ such that $0\in\left(  D\right)  ^{N}\cap\mathfrak{D}%
\neq\mathfrak{\varnothing,}$ where $N=\dim_{\mathbb{C}}$ $\mathfrak{M}_{L}%
($M$)\mathfrak{.}$ To prove Theorem \ref{Nik} we need to derive an explicit
formula for the metric $\left\langle \omega_{\tau},\omega_{\tau}\right\rangle
:=h(\tau,\overline{\tau})$ on the line bundle $\pi_{\ast}\left(
\omega_{\mathcal{X}\text{/}\mathfrak{M}_{L}(\text{M})}\right)  $ restricted
on
\[
\left(  D\right)  ^{N}-\left(  \left(  D\right)  ^{N}\cap\mathfrak{D}\right)
=\left(  D^{\ast}\right)  ^{k}\times\left(  D\right)  ^{N-k}.
\]
Let (\textrm{U}$^{i})^{k}\times(D)^{N-k}$ be the universal cover of
\[
\left(  D\right)  ^{N}-\left(  \left(  D\right)  ^{N}\cap\mathfrak{D}\right)
=\left(  D^{\ast}\right)  ^{k}\times\left(  D\right)  ^{N-k},
\]
where \textrm{U}$^{i}$ are the unit disks. Let%
\begin{equation}
\pi:(\mathrm{U}^{i})^{k}\times(D)^{N-k}\rightarrow\left(  D^{\ast}\right)
^{k}\times\left(  D\right)  ^{N-k} \label{Un}%
\end{equation}
be the covering map.

We will prove that formula $\left(  \ref{form}\right)  $ implies that we have
the following expression for the $L^{2}$ metric $\left\langle \omega_{\tau
},\omega_{\tau}\right\rangle :=h(\tau,\overline{\tau})$ on $\pi_{\ast}\left(
\omega_{\mathcal{X}\text{/}\mathfrak{M}_{L}(\text{M})}\right)  $ restricted on
the universal covering of (\textrm{U}$^{i})^{k}\times(D)^{N-k}$ of $\left(
D^{\ast}\right)  ^{k}\times\left(  D\right)  ^{N-k}:$%
\[
\left\langle \omega_{\tau},\omega_{\tau}\right\rangle :=h(\tau,\overline{\tau
}):=
\]
\begin{equation}%
{\displaystyle\sum\limits_{i=1}^{k}}
\left(  1-|\tau^{i}|^{2}\right)  +%
{\displaystyle\sum\limits_{j=k+1}^{N}}
\left(  1-|t^{j}|^{2}\right)  +\phi(\tau,\overline{\tau})+\Psi(t,\overline{t})
\label{h}%
\end{equation}
where $\phi(\tau,\overline{\tau}),$ $\Psi(t,\overline{t})$ are bounded real
analytic functions on (\textrm{U}$^{i})^{k}\times(D)^{N-k}.$\ Theorem
\ref{gmet} and formula $\left(  \ref{h}\right)  $ imply Theorem \ref{Nik}.

\textbf{Proof}: The proof will follow from the Lemma proved bellow.

\begin{lemma}
\label{LIM1}Let $\left(  \mathrm{U}\right)  ^{k}$ be the universal cover of
$(D^{\ast})^{k}:=\left(  D\right)  ^{k}-\left(  D\right)  ^{k}\cap
\mathfrak{M}_{L}($M$)$ where $\mathrm{U}$ is the unit disk.  Then
\end{lemma}

\textbf{i. }\textit{There exists a family of CY manifolds }%
\begin{equation}
\pi^{\ast}\left(  \mathcal{X}_{\left(  \mathrm{U}\right)  ^{k}}\right)
\rightarrow\left(  \mathrm{U}\right)  ^{k}. \label{Fam}%
\end{equation}
\textit{over }$\left(  \mathrm{U}\right)  ^{k}$ \textit{and a holomorphic
section }$\omega\in H^{0}\left(  \left(  \mathrm{U}\right)  ^{k},\pi_{\ast
}\omega_{\mathcal{X}\text{(M)/}\left(  \mathrm{U}\right)  ^{k}}\right)  $
\textit{such that }$\omega_{\tau}=\omega|_{\text{M}_{\tau}}$\textit{ is a non
zero holomorphic form on M}$_{\tau}.$\textit{ }

\textbf{ii. }$\left\langle \omega_{\tau},\omega_{\tau}\right\rangle $
\textit{can be represented on }$\left(  \mathrm{U}\right)  ^{k}$ \textit{as
follows}$:$%
\begin{equation}
\left\langle \lbrack\omega_{\tau}],[\omega_{\tau}]\right\rangle =h(\tau
,\overline{\tau})=%
{\displaystyle\sum\limits_{i=1}^{k}}
\left(  1-\left\vert \tau^{i}\right\vert ^{2}\right)  +\phi(\tau
,\overline{\tau}),\label{log2}%
\end{equation}
\textit{where }$\phi(\tau,\overline{\tau})$\textit{ is a bounded real analytic
functions on} $\left(  \mathrm{U}\right)  ^{k}$ \textit{such that} \textit{the
limits}%
\[
\underset{\tau\rightarrow\mathfrak{\kappa}_{\infty}\in\left(  \overline
{\mathrm{U}}\right)  ^{k}-(\mathrm{U})^{k}}{\lim}h(\tau,\overline{\tau})\text{
and }\underset{\tau\rightarrow\mathfrak{\kappa}_{\infty}\in\left(
\overline{\mathrm{U}}\right)  ^{k}-(\mathrm{U})^{k}}{\lim}\phi(\tau
,\overline{\tau})
\]
\textit{exist where}%
\[
\mathfrak{\kappa}_{\infty}=\left(  \kappa_{\infty}^{1},...,\kappa_{\infty}%
^{k}\right)  =
\]%
\[
\underset{(q^{1},...,q^{k})\rightarrow(0,...,0)}{\lim}\left(  \pi^{-1}%
(q^{1})=\tau^{1},...,\pi^{-1}(q^{1})=\tau^{k}\right)  ,
\]%
\[
(q^{1},...,q^{k})\in(D^{\ast})^{k},\text{ }\pi\left(  \mathfrak{\kappa
}_{\infty}\right)  =(0,...,0)\in\left(  \overline{D^{\ast}}\right)  ^{k}=D^{k}%
\]
\textit{and }$\left(  \kappa_{\infty}^{1},...,\kappa_{\infty}^{k}\right)
\in\left(  \partial\mathrm{U}\right)  ^{k}.$

\textbf{iii.} \textit{Suppose that }$\underset{\tau\rightarrow\mathfrak{\kappa
}_{\infty}\in\left(  \overline{\mathrm{U}}\right)  ^{k}-(\mathrm{U})^{k}}%
{\lim}h(\tau,\overline{\tau})=0.$ \textit{Then }%
\[
\underset{\tau\rightarrow\mathfrak{\kappa}_{\infty}\in\left(  \overline
{\mathrm{U}}\right)  ^{k}-(\mathrm{U})^{k}}{\lim}\phi(\tau,\overline{\tau})=0.
\]

\textbf{iv}\textit{. The function }$\phi(\tau,\overline{\tau})$\textit{ is
bounded on }$\left(  \overline{\mathrm{U}}\right)  ^{k}.$

\textbf{Proof of i: }Since $(D^{\ast})^{k}\subset\mathfrak{M}_{L}$(M), Remark
\ref{Vie1} implies that there exists a family of CY manifolds
\begin{equation}
\pi:\mathcal{X}_{(D^{\ast})^{k}}\rightarrow(D^{\ast})^{k} \label{F0}%
\end{equation}
over $(D^{\ast})^{k}.$ Because $p:$ $(\mathrm{U})^{k}\rightarrow(D^{\ast}%
)^{k}$ is the universal cover of $(D^{\ast})^{k},$ then the pull back of the
family $\left(  \ref{F0}\right)  $ by $p$ defines the family $\left(
\ref{Fam}\right)  .$ Let $\omega_{\tau}$ be the section of the pullback of the
restriction of the relative dualizing line bundle $\omega_{\mathcal{X}%
/\mathfrak{M}_{L}\text{(M)}}$ on $(D^{\ast})^{k}$ on $(\mathrm{U})^{k}$
constructed in Theorem \ref{FCS}. $\blacksquare$\textbf{ }

\textbf{Proof of ii: }The proof of Part \textbf{ii }is based on the following Proposition:

\begin{proposition}
\label{lim}Let us consider $\left(  D_{\alpha_{1},\alpha_{2}}\right)
^{k}\subset(D^{\ast})^{k}\subset(D)^{k}\subset\mathfrak{M}_{L}($M$),$ where%
\[
D_{\alpha_{1},\alpha_{2}}:=\left\{  t\in D_{\alpha_{1},\alpha_{2}}\left\vert
\left\vert t\right\vert <1\text{ and }\alpha_{1}<\arg t<\alpha_{2}\right.
\right\}  .
\]
Suppose the closure of $\left(  D_{\alpha_{1},\alpha_{2}}\right)  ^{k}$ in
$(D)^{k}$ contains $0\in\left(  \overline{D^{\ast}}\right)  ^{k}=D^{k}.$ Let
us consider the restriction of the family $\left(  \ref{Fam}\right)  $%
\begin{equation}
\mathcal{X}_{\alpha_{1},\alpha_{2}}\rightarrow\left(  D_{\alpha_{1},\alpha
_{2}}\right)  ^{k}\label{famr}%
\end{equation}
on $\left(  D_{\alpha_{1},\alpha_{2}}\right)  ^{k}\subset\mathfrak{M}_{L}%
($M$).$ Let $\omega_{\mathcal{X}_{\alpha_{1},\alpha_{2}}\left/  \left(
D_{\alpha_{1},\alpha_{2}}\right)  ^{k}\right.  }$ be the restriction of
dualizing sheaf of the family of polarized CY manifolds $\left(
\ref{Fam}\right)  $ on $\left(  D_{\alpha_{1},\alpha_{2}}\right)  ^{k}$. Then
there exists a global section
\[
\eta\in\Gamma\left(  \left(  D_{\alpha_{0},\alpha_{1}}\right)  ^{k},\pi_{\ast
}\omega_{\mathcal{X}_{\alpha_{1},\alpha_{2}}\left/  \left(  D_{\alpha
_{1},\alpha_{2}}\right)  ^{k}\right.  }\right)
\]
such that the classes of cohomology $[\eta_{q}]$ defined by the restriction of
$\eta$ on all of the fibres $\pi^{-1}(q):=$M$_{q}$ for $q\in\left(
D_{\alpha_{1},\alpha_{2}}\right)  ^{k}$ are non zero elements of $H^{0}%
($M$_{q},\Omega_{\text{M}_{q}}^{n}).$ The limit $\underset{q\rightarrow0}%
{\lim}[\eta_{q}]$ exists and
\begin{equation}
\underset{q\rightarrow0}{\lim}[\eta_{q}]=[\eta_{0}]\text{ and }\left\langle
[\eta_{0}],[\eta_{0}]\right\rangle \geq0.\label{lim0}%
\end{equation}

\end{proposition}

\textbf{Proof: }$\left(  D_{\alpha_{1},\alpha_{2}}\right)  ^{k}$ is a
contractible sector in $\left(  D^{\ast}\right)  ^{k}.$ Thus if we fix a basis
$(\gamma_{1},...,\gamma_{b_{n}})$ in $H^{n}($M,$\mathbb{Z)}/Tor$ then we are
fixing the marking of the family $\left(  \ref{famr}\right)  $ over each point
M$_{\tau}$ for each point $\tau\in\left(  D_{\alpha_{1},\alpha_{2}}\right)
^{k}.$ This means that the basis $(\gamma_{1},...,\gamma_{b_{n}})$ of $H^{n}%
($M,$\mathbb{Z)}/Tor$ is defined and fixed on $H^{n}($M$_{\tau}$%
,$\mathbb{Z)}/Tor$ on each fibre of the family $\left(  \ref{famr}\right)  $.
Now \ we can define the period map of the family by
\[
p(q):=\left(  ...,%
{\displaystyle\int\limits_{\gamma_{i}}}
\eta_{q},...\right)  ,
\]
where $\eta_{q}$ is a non zero holomorphic form on $\pi^{-1}(q):=$M$_{q}.$
Local Torelli Theorem implies that the period map $p$ of marked CY manifolds%
\begin{equation}
p:\left(  D_{\alpha_{1},\alpha_{2}}\right)  ^{k}\rightarrow\mathbb{P}%
(H^{n}(\text{M},\mathbb{C})).\label{incl}%
\end{equation}
is an embedding $\left(  D_{\alpha_{1},\alpha_{2}}\right)  ^{k}\subset
\mathbb{P}(H^{n}($M$,\mathbb{C})).$ Thus we can conclude from the compactness
of $\mathbb{P}(H^{n}($M$,\mathbb{C}))$ and $\left(  \ref{incl}\right)  $ the
existence of a sequence of $[\eta_{q}]$ such that
\begin{equation}
\underset{q\rightarrow0}{\lim}[\eta_{q}]=[\eta_{0}]\label{limb}%
\end{equation}
exists and $\eta|_{\text{M}_{0}}\neq0.$ $\left(  \ref{limb}\right)  $
\ implies $\left(  \ref{lim0}\right)  .$ $\blacksquare$

\begin{corollary}
\label{lim0a}There exists a global section $\eta\in H^{0}\left(
(\mathrm{U})^{k},\omega_{\mathcal{X}_{(\mathrm{U})^{k}}\left/  (\mathrm{U}%
)^{k}\right.  }\right)  $ such that $\underset{\tau\rightarrow\kappa_{\infty}%
}{\lim}[\eta_{\tau}]$ exists and%
\begin{equation}
\underset{\tau\rightarrow\kappa_{\infty}}{\lim}[\eta_{\tau}]=[\eta
_{\kappa_{\infty}}]\neq0. \label{limc}%
\end{equation}

\end{corollary}

\begin{proposition}
\label{LIM3}Let $\left\{  \omega_{\tau}\right\}  $ be the family of
holomorphic $n-$forms constructed in Theorem \ref{FCS} on the family
restricted on $(\mathrm{U})^{k}.$ Then the limit $\underset{\tau
\rightarrow\kappa_{\infty}\in\left(  \overline{\mathrm{U}}\right)
^{k}-(\mathrm{U})^{k}}{\lim}[\omega_{\tau}]$ exists,
\begin{equation}
\underset{\tau\rightarrow\kappa_{\infty}\in\left(  \overline{\mathrm{U}%
}\right)  ^{k}-(\mathrm{U})^{k}}{\lim}=\left[  \omega_{\infty}\right]  \text{
and }\left\langle \left[  \omega_{\infty}\right]  ,\left[  \omega_{\infty
}\right]  \right\rangle \geq0. \label{fam5a}%
\end{equation}

\end{proposition}

\textbf{Proof:} According to Corollary \ref{lim0a} there exists a global
section
\[
\eta\in H^{0}\left(  (\mathrm{U})^{k},\omega_{\mathcal{X}_{(\mathrm{U})^{k}%
}\left/  (\mathrm{U})^{k}\right.  }\right)
\]
such that $\underset{\tau\rightarrow\kappa_{\infty}}{\lim}[\eta_{\tau}]$
$\ $satisfies $\left(  \ref{limc}\right)  .$ The relation between the
cohomologies of holomorphic forms $\eta_{\tau}:=\eta_{q}$ and $\omega_{\tau}$
are given by the formula $[\eta_{\tau}]=\varphi(\tau)[\omega_{\tau}],$ where
$\varphi(\tau)$ is a holomorphic function on the product $(\mathrm{U})^{k}.$
According to Theorem \ref{forms} we have
\begin{equation}
0\leq\left\langle \lbrack\omega_{\tau}],[\omega_{\tau}]\right\rangle
\leq\left\langle \lbrack\omega_{\tau_{0}}],[\omega_{\tau_{0}}]\right\rangle
.\label{ineq}%
\end{equation}
Thus $\left(  \ref{incl}\right)  ,$ $\left(  \ref{ineq}\right)  $\ and
$[\eta_{\tau}]=\varphi(\tau)[\omega_{\tau}]$ imply formula $\left(
\ref{fam5a}\right)  $. So the limit
\[
\underset{\tau\rightarrow\kappa_{\infty}\in\left(  \overline{\mathrm{U}%
}\right)  ^{k}-(\mathrm{U)}^{k}}{\lim}\left\langle [\omega_{\tau}%
],[\omega_{\tau}]\right\rangle
\]
exists and
\[
\underset{\tau\rightarrow\kappa_{\infty}\in\left(  \overline{\mathrm{U}%
}\right)  ^{k}-(\mathrm{U)}^{k}}{\lim}\left\langle [\omega_{\tau}%
],[\omega_{\tau}]\right\rangle =\underset{\tau\rightarrow\kappa_{\infty}%
\in\left(  \overline{\mathrm{U}}\right)  ^{k}-(\mathrm{U)}^{k}}{\lim
h(\tau,\overline{\tau})}=h(\kappa_{\infty})\geq0.
\]
Proposition \ref{LIM3} is proved. $\blacksquare$

\begin{corollary}
\label{LIM2}Let $\left[  \omega_{\infty}\right]  $ be defined by $\left(
\ref{fam5a}\right)  .$ Then $\left\langle \left[  \omega_{\infty}\right]
,\left[  \omega_{\infty}\right]  \right\rangle =0$ if and only if the
monodromy of the restriction of the family $\left(  \ref{FAM}\right)  $ is infinite.
\end{corollary}

Notice that the functions $\left\langle \left\langle [\eta_{\tau}],[\eta
_{\tau}]\right\rangle \right\rangle $ and $\left\langle [\omega_{\tau
}],[\omega_{\tau}]\right\rangle $ are real analytic. If we normalize
$\omega_{0}$ and $\phi_{i}$ such that
\[
\left\Vert \omega_{\tau_{0}}\right\Vert ^{2}=\left\langle \omega_{0}%
,\omega_{0}\right\rangle =1
\]
and
\[
\left\langle \omega_{0}\lrcorner\phi_{i},\omega_{0}\lrcorner\phi
_{j}\right\rangle =\delta_{i\overline{j}}%
\]
we get from $\left(  \ref{form}\right)  $ the following expression%
\[
h(\tau,\overline{\tau})=
\]%
\[
1-%
{\displaystyle\sum\limits_{i=1}^{k}}
|\tau^{i}|^{2}+%
{\displaystyle\sum\limits_{i\leq j}}
\left\langle \omega_{0}\lrcorner\left(  \phi_{i}\wedge\phi_{k}\right)
,\omega_{0}\lrcorner\left(  \phi_{j}\wedge\phi_{l}\right)  \right\rangle
\tau^{i}\overline{\tau^{j}}\tau^{k}\overline{\tau^{l}}+O(\tau^{5})=
\]%
\begin{equation}
1-%
{\displaystyle\sum\limits_{i=1}^{k}}
|\tau^{i}|^{2}+\Phi(\tau,\overline{\tau}) \label{log11}%
\end{equation}
holds. Also $\left(  \ref{log11}\right)  $ implies that the restriction of
$\left\langle [\omega_{\tau}],[\omega_{\tau}]\right\rangle =h(\tau
,\overline{\tau})$\ on the universal cover ($\mathrm{U)}^{k}$of ($D^{\ast
}\mathrm{)}^{k}$ will be given by $\left(  \ref{log2}\right)  $, i.e.
\[
\left\langle \lbrack\omega_{\tau}],[\omega_{\tau}]\right\rangle =h(\tau
,\overline{\tau})=1-%
{\displaystyle\sum\limits_{i=1}^{k}}
\left\vert \tau^{i}\right\vert ^{2}+\Phi(\tau,\overline{\tau}).
\]
\textbf{Proof of }$\left(  \ref{log2}\right)  :$ We can rewrite the above
expression as follows:%
\[
\left\langle \lbrack\omega_{\tau}],[\omega_{\tau}]\right\rangle =h(\tau
,\overline{\tau})=1-%
{\displaystyle\sum\limits_{i=1}^{k}}
\left\vert \tau^{i}\right\vert ^{2}+\Phi(\tau,\overline{\tau})=
\]%
\begin{equation}%
{\displaystyle\sum\limits_{i=1}^{k}}
\left(  1-\left\vert \tau^{i}\right\vert ^{2}\right)  -k+1+\Phi(\tau
,\overline{\tau})=%
{\displaystyle\sum\limits_{i=1}^{k}}
\left(  1-\left\vert \tau^{i}\right\vert ^{2}\right)  +\phi(\tau
,\overline{\tau}). \label{log3}%
\end{equation}
where $\phi(\tau,\overline{\tau})=\Phi(\tau,\overline{\tau})-k+1.$ Proposition
\ref{LIM3} and $\left(  \ref{fam5a}\right)  $ imply that%
\[
\underset{\tau\rightarrow\kappa_{\infty}\in\left(  \overline{\mathrm{U}%
}\right)  ^{k}-(\mathrm{U)}^{k}}{\lim}h(\tau,\overline{\tau}),\text{
}\underset{\tau\rightarrow\kappa_{\infty}\in\left(  \overline{\mathrm{U}%
}\right)  ^{k}-(\mathrm{U)}^{k}}{\lim}\phi(\tau,\overline{\tau})
\]
exist and $\phi(\tau,\overline{\tau})$ is a bounded real analytic function on
($\mathrm{U)}^{k}.$ Part \textbf{ii }of Lemma \ref{LIM1} is proved.
$\blacksquare$

\textbf{Proof of part iii:} Suppose that $\mathrm{U}$ is the unit disk and
\[
\underset{\tau\rightarrow\kappa_{\infty}\in\left(  \overline{\mathrm{U}%
}\right)  ^{k}-(\mathrm{U)}^{k}}{\lim}\left\langle \omega_{\tau},\omega_{\tau
}\right\rangle =\text{ }\underset{\tau\rightarrow\kappa_{\infty}\in\left(
\overline{\mathrm{U}}\right)  ^{k}-(\mathrm{U)}^{k}}{\lim}h(\tau
,\overline{\tau})=0.\text{ }%
\]
Notice that since $\kappa_{\infty}\in\left(  \overline{\mathrm{U}}\right)
^{k}-(\mathrm{U)}^{k},$ where $\mathrm{U}$ is the unit disk then for each $i$
we have $\left(  1-\left\vert \kappa_{\infty}^{i}\right\vert ^{2}\right)  =0$
and thus
\begin{equation}%
{\displaystyle\sum\limits_{i=1}^{k}}
\left(  1-\left\vert \kappa_{\infty}^{i}\right\vert ^{2}\right)
=0.\label{log3a}%
\end{equation}
Thus $\left(  \ref{log3}\right)  $ and $\left(  \ref{log3a}\right)  $ imply
$\left(  \ref{log2}\right)  .$ Part \textbf{iii} is proved. $\blacksquare$

Part \textbf{iii} implies Part \textbf{iv. }$\blacksquare$

Lemma \ref{LIM1} is proved. $\blacksquare$

\begin{lemma}
\label{1}Suppose that the $L^{2}$ metric on the relative dualizing sheaf
defined by the function $h(\tau,\overline{\tau})=\left\langle \omega_{\tau
},\omega_{\tau}\right\rangle $ on
\[
D^{N}-\left(  D^{N}\cap\mathfrak{D}\right)  =(D^{\ast})^{k}\times
D^{N-k}\subset\mathfrak{M}_{L}(\text{M}).
\]
is bounded on $D^{N},$ $h|_{(D)^{N}-\left(  D\right)  ^{N}\cap\mathfrak{D}}>0$
and $h|_{\left(  D\right)  ^{N}\cap\mathfrak{D}}\geq0.$ Then the $L^{2}$
metric is good.
\end{lemma}

\textbf{Proof: }The proof of Lemma \ref{1} is obvious. $\blacksquare$

The expression $\left(  \ref{log2}\right)  $ for the $L^{2}$ metric and
Theorem \ref{gmet} implies that if
\[
h|_{\left(  D\right)  ^{N}\cap\mathfrak{D}}\geq0
\]
then the $L^{2}$ metric is a good metric . Theorem \ref{Nik} is proved.
$\blacksquare$

\subsection{The Weil-Petersson Volumes are Rational Numbers}

\begin{theorem}
\label{vol}The Weil-Petersson volume of the moduli space of polarized CY
manifolds is finite and it is a rational number.
\end{theorem}

\textbf{Proof: }Theorem \ref{Nik} implies that the metric on the relative
dualizing sheaf $\omega_{\mathcal{X}/\mathfrak{M}_{L}(\text{M)}}$ defined by
$\left(  \ref{W-P}\right)  $ is a good metric. This implies that the Chern
form of any good metric defines a class of cohomology in
\[
H^{2}\left(  \overline{\mathfrak{M}_{L}(\text{M})},\mathbb{Z}\right)  \cap
H^{1,1}\left(  \overline{\mathfrak{M}_{L}(\text{M})},\mathbb{Z}\right)  .
\]
See Theorem \ref{Mum100}. We know from \cite{To89} that the Chern form of the
metric $h$ is equal to minus the imaginary part of the Weil-Petersson metric.
So the imaginary part of the Weil-Petersson metric is a good form in the sense
of Mumford. This implies that%
\[%
{\displaystyle\int\limits_{\overline{\mathfrak{M}_{L}(\text{M})}}}
\wedge^{\dim_{\mathbb{C}}\overline{\mathfrak{M}_{L}(\text{M})}}c_{1}%
(h)\in\mathbb{Z}%
\]
since $\overline{\mathfrak{M}_{L}(\text{M})}$ is a smooth manifold. Since
$\mathfrak{M}_{L}($M$)$ is a finite cover of the moduli space $\mathcal{M}%
_{L}($M$)\,$then the Weil-Petersson volume of $\mathcal{M}_{L}($M$)$ will be a
rational number. Theorem \ref{vol} is proved. $\blacksquare$

In the paper \cite{LS1} the authors proved that the Weil-Petersson volumes of
the moduli space of CY manifolds are finite.

\begin{corollary}
\label{vol1}The Weil-Petersson metric is a good metric on the moduli space
$\overline{\mathfrak{M}_{L}(\text{M})}$ and the Chern forms $c_{k}\left[
W.-P.\right]  $ of the Weil-Petersson metric are well defined elements of
$H^{2k}\left(  \overline{\mathfrak{M}_{L}(\text{M})},\mathbb{Z}\right)  .$
\end{corollary}

\end{document}